\documentclass[preprint,aps,floats,prb,fancyhdr,a4paper,superscriptaddress,nofootinbib,showpacs]{revtex4}
\usepackage{latexsym}
\usepackage{amsmath}
\usepackage{amssymb}
\usepackage{amsfonts}
\usepackage{graphicx}
\usepackage{dcolumn}
\usepackage{fancyhdr}
\usepackage[colorlinks]{hyperref}

\newtheorem{theorem}{Theorem}
\newtheorem{acknowledgement}[theorem]{Acknowledgement}

\begin{document}

\title{Cosmological Models with Variable Constants.\\
Their Solution Through Similarity Methods.}
\author{Jos\'{e} Antonio Belinch\'{o}n \& Pedro D\'{a}vila \\
Dept. F\'{\i}sica ETS Arquitectura UPM\\
Av. Juan de Herrera 4 Madrid 28040 Espa\~{n}a}
\date{}

\begin{abstract}
\emph{In this work we compile a few differential equations (ODEs) that arise
from the relativistic equations in cosmological models that consider the
``constants'' as scalars functions dependent on time and they are described
as perfect as well as viscous fluids. The general idea of the paper is to
show how to solve the equations of the models through dimensional techniques
(self-similarity). The results are compared with those obtained by other
authors and new solutions are introduced.}
\end{abstract}

\maketitle

\section{\textbf{Introduction.}}

The purpose of this work is to study different cosmological models that
envisage the classical constant as scalar functions dependent on time. To
outline the differential equations that govern such models and to compare
their solution through the traditional method (integration of ODEs, this is
the way followed by other authors) with the Dimensional Method
(self-similarity). The latter method will consist in the reduction of the
number of variables and the obtainment of ODEs easily integrable (see \cite
{B},\cite{BAR} and in special \cite{CAR} for a review of self-similarity in
General Relativity)\medskip\

In all the studied cases, the equations are very similar, there will only be
differences when the term $p$ (pressure) is defined, depending on if it is
considered a perfect or viscous fluid. The modified field equations are:
(the ``constants'' $G,c$ and $\Lambda $ are functions on $t$ ).
\begin{equation}
R_{ij}-\frac{1}{2}g_{ij}R-\Lambda g_{ij}=\frac{8\pi G}{c^{4}}T_{ij}
\label{eg1}
\end{equation}
where $\Lambda $ represents the cosmological ``constant'' and we impose that
the second Bianchi identity is verified: (after raising an index)
\begin{equation}
\left( R_{i}^{j}-\frac{1}{2}\delta _{i}^{j}R\right) _{;j}=\left( \frac{8\pi G%
}{c^{4}}T_{i}^{j}+\Lambda \delta _{i}^{j}\right) _{;j}  \label{eg2}
\end{equation}
as well as the so-called conservation principle (bad so-called\footnote{%
we understand that this is only an expresion since in GR the conservation of
the energy and momentum is only approximate} see \cite{LAND}) for the
energy-momentum tensor i.e. the covariant divergence of the stress-energy
tensor:
\begin{equation}
div(T_{i}^{j})=0  \label{eg3}
\end{equation}
this condition will be considered in some cases since we shall see we that
can avoid it.

\begin{enumerate}
\item  The line element is defined by:
\begin{equation*}
ds^{2}=-c^{2}dt^{2}+f^{2}(t)\left[ \frac{dr^{2}}{1-kr^{2}}+r^{2}\left(
d\theta ^{2}+\sin {}^{2}\theta d\phi ^{2}\right) \right]
\end{equation*}
we only consider the case $k=0,$ here$.$

\item  The energy-momentum tensor is defined by:
\begin{equation*}
T_{ij}=(\rho +p)u_{i}u_{j}-pg_{ij}
\end{equation*}
where $\rho $ is the energy density and $p$ represents pressure (to be
defined in a generical way), always verifying $\left[ \rho \right] =\left[ p%
\right] $.
\end{enumerate}

The paper is organized as follows: In the second section, a model will be
studied that only envisages the ``constants'' $G$ and $\Lambda $ as variable
and whose energy-momentum tensor is described by a perfect fluid. As in all
the studied cases, this section will begin outlining the equations that
govern such model, going next to study its detailed solution through the two
techniques commented previously. The traditional one, will consist in the
math integration of the equations (non dimensional method). The other is
based on dimensional techniques. The latter will be studied in two ways, the
first one, that we shall designate as the ``\emph{simplest method'' (naive)}%
, will consist in the dimensional study of the problem (model) analyzing its
set of governing quantities. This set will lead us to the solution of the
outlined equations through the application of the Pi theorem. While the
other dimensional technique that we shall study and that we designate as
\emph{not so simple method}, is a mixture of the classic dimensional
analysis (obtainment of $\pi $-monomials) with the direct integration of the
equations. The pi monomials will \emph{reduce the number of variables} in
the differential equations and therefore it will bring us to obtain a
simpler differential equation (ODEs) that will be directly integrated. The
efficiency of Dimensional Analysis will enable us to provide new solutions
even in the cases in which we do not envisage some of the departure
hypotheses as for example $div(T_{i}^{j})=0$. In the third section we shall
study a model that envisages $G$ as well as $\Lambda $ as scalar functions
and whose tensor is characterized by a bulk viscous fluid. As in the second
paragraph, the classic solution of the problem will be studied and it will
be compared to the one obtained through the dimensional techniques,
providing new solutions in this case. In the fourth section, constant $G$, $%
c $ and $\Lambda $ are envisaged as dependent functions on time within a
model described by a bulk viscous fluid. Only in this case they are provided
dimensional solutions since, at the moment, they are the only ones known. In
the fifth section a model will be studied that may be understood as a
generalization of the previous one upon considering in it mechanisms of
adiabatic creation of matter. In this case, it is observed that the
governing equations of the model are reduced to the case above therefore no
further commentaries are needed. Finally it will be ended with some brief
conclusions.

\section{\textbf{Case $G$ and $\Lambda $ variable for a perfect fluid.}}

Attending to the specifications made in the introduction in this case $p$
(pressure) verifies the following state equation:
\begin{equation}
p=\omega \rho \qquad \omega =const.  \label{ee1}
\end{equation}
where $\omega \in \left[ 0,1\right] $ (i.e. it is a pure number) so that the
energy-momentum tensor verifies the energy conditions. Under these
circumstances the equations are:
\begin{equation}
2\frac{f\,^{\prime \prime }}{f\,}+\frac{(f\,^{\prime })^{2}}{f\,^{2}}=-\frac{%
8\pi G(t)}{c^{2}}p+c^{2}\Lambda (t)  \label{p1}
\end{equation}
\begin{equation}
3\frac{(f\,^{\prime })^{2}}{f\,^{2}}=\frac{8\pi G(t)}{\,c^{2}}\rho
+c^{2}\Lambda (t)  \label{p2}
\end{equation}
from the expressions (\ref{p1}) and (\ref{p2}) we obtain the following
relationship
\begin{equation}
G\rho ^{\prime }+3(1+\omega )\rho G\frac{f^{\prime }}{f}+\rho G^{\prime }+%
\frac{\Lambda ^{\prime }c^{4}}{8\pi }=0
\end{equation}
furthermore the following law is taken into account
\begin{equation}
div(T_{i}^{j})=0\text{ }\Leftrightarrow \rho ^{\prime }+3(\rho +p)\frac{%
f^{\prime }}{f}=0  \label{p3}
\end{equation}
Now, we go on to see how this model is solved through two methods, one of
them, by traditional integration and the other by dimensional approach.

\subsection{\textbf{Non Dimensional method.}}

Deriving the equation (\ref{p2}) and simplifying with (\ref{p1}) we obtain
the following relationship. This relationship was used by Lau who reconciled
the LNH of Dirac with the GR (see \cite{LAU})
\begin{equation}
G\rho ^{\prime }+3(1+\omega )\rho G\frac{f^{\prime }}{f}+\rho G^{\prime }+%
\frac{\Lambda ^{\prime }c^{4}}{8\pi }=0  \label{f1}
\end{equation}
regrouping terms and taking into account $div(T_{i}^{j})=0$ we obtain:
\begin{equation}
\rho ^{\prime }+3(\rho +p)\frac{f^{\prime }}{f}=-\left[ \frac{\Lambda
^{\prime }c^{4}}{8\pi G}+\rho \frac{G^{\prime }}{G}\right]  \label{SIL0}
\end{equation}

From the equations (\ref{f1}) and (\ref{p3}) the following equation that
relates $G$ with $\Lambda $ is obtained.
\begin{equation}
G^{\prime }=-\frac{\Lambda ^{\prime }c^{4}}{8\pi \rho }  \label{f2}
\end{equation}
From all these relationships the following differential equation is obtained
which is not immediately integrable: combining (\ref{p2})) with (\ref{p3})
(here we are continuing Kalligas et al`s work see \cite{KA})
\begin{equation*}
\left( \frac{\rho ^{\prime }}{\rho }\right) ^{2}=9(1+\omega )^{2}\left(
\frac{8\pi G}{3c^{2}}\rho +\frac{\Lambda c^{2}}{3}\right)
\end{equation*}
deriving with respect to $t$ then we obtain:
\begin{equation}
\frac{\rho ^{\prime }\rho ^{\prime \prime }}{\rho ^{2}}-\left( \frac{\rho
^{\prime }}{\rho }\right) ^{3}=12\pi (\omega +1)^{2}\frac{G\rho ^{\prime }}{%
c^{2}}  \label{f4}
\end{equation}
that is to say:
\begin{equation}
\rho \rho ^{\prime \prime }-\left( \rho ^{\prime }\right) ^{2}=12\pi (\omega
+1)^{2}\frac{G\rho ^{3}}{c^{2}}  \label{f5}
\end{equation}
To integrate this equation Kalligas et al made the following hypothesis on
the behavior of the function $G:$ $G\propto t^{\alpha }$ with $\alpha \in
\mathbb{R}$ (see \cite{KA}) this is an unfounded hypothesis in our opinion
leading to $G=Ct^{\alpha }$ (where $C$ represents certain constant of
proportionality). The obtained results are:
\begin{equation*}
\rho (t)=\dfrac{\alpha +2}{12\pi (\omega +1)^{2}C}\dfrac{1}{t^{\alpha +2}}
\end{equation*}
etc... (see \cite{KA}). We believe that this solution, even though correct,
under the outlined hypotheses presents certain degree of arbitrariness,
since we will know neither the behavior of $G$ nor the one of any other
quantity in function of the equation of state that we are imposing i.e. we
should give any value to $\alpha $ and ``\emph{guess''} which model is
describing i.e. what state equation belongs to, furthermore they introduce a
new dimensional constant, $C,$ of doubtful physical meaning.

\subsection{\textbf{Dimensional Method.}}

In this section, two dimensional tactics are studied. The first of them,
designated as ``\emph{naive method''} (since it is always the simplest
method as well as effective) studies the set of governing quantities of the
problem and through the application of the Pi theorem a solution to the
equations except the numerical constant ones is obtained. The second method,
the one designated as ``\emph{not so simplest method''}, combines the
classic dimensional analysis with the direct integration of the differential
equations. This method has been developed by Prof. M. Casta\~{n}s for
ordinary differential equations (see \cite{CAS1} and \cite{CAS2})

\subsubsection{\textbf{Naive Method.}}

With the dimensional method that we have followed we are not forced to
impose similar condition and obviously the results that we obtain are as
good or even better since our results depend on the state equation that is
imposed (see \cite{T1} and \cite{T2}). The trick is based in integrating the
equation $\rho ^{\prime }+3(\rho +p)\frac{f^{\prime }}{f}=0$ from which we
obtain a dimensional constant $A_{\omega }$ indispensable for our trifling
count. With this constant the set of governing parameters is: $\frak{M=M}%
(t,c,A_{\omega })$. With these quantities in a dimensional base $\frak{B}%
=\left\{ L,M,T\right\} $ the problem remains perfectly determined with not
further conditions (see \cite{T1} and \cite{T2}).

\subsubsection{\textbf{Not so simple Method}}

In this section we shall study several of the possibilities that may arise.

\paragraph{\textbf{Considering $div(T_{i}^{j})=0$}$.$}

We note in the equation
\begin{equation}
\underset{A1}{\underbrace{\rho ^{\prime }+3(\omega +1)\rho \frac{f^{\prime }%
}{f}}}=-\underset{A2}{\underbrace{\left[ \frac{\Lambda ^{\prime }c^{4}}{8\pi
G}+\rho \frac{G^{\prime }}{G}\right] }}
\end{equation}
if the conservation principle for the energy-momentum tensor is taken into
account, from the part $\left( A1\right) $ of the equation we obtain the
well-known relationship $\ \rho =A_{\omega }f^{-3(\omega +1)}.$ Regarding to
the second term of the equation $\left( A2\right) $ from it, we can extract
a $\pi $-monomial $\pi _{1}=\Lambda c^{2}t^{2}$ that we may express through
the following equality $\Lambda =\frac{d}{c^{2}t^{2}}$ where $d\in \mathbb{R}
$ i.e. it is a pure number. This $\pi $-monomial is replaced into the
equation in the following way:
\begin{equation}
\frac{dc^{2}}{4\pi t^{3}}=\frac{A_{\omega }G^{\prime }}{f^{3(\omega +1)}}
\end{equation}
From this equation we can obtain another $\pi $-monomial, $\pi
_{2}=fc^{-1}t^{-1}$ that we may express as $f=act$ where $a\in \mathbb{R}$
is a numerical constant. With this new relationship we simplify our last
equation, yielding:
\begin{equation*}
\frac{dc^{2}}{4\pi t^{3}}=\frac{A_{\omega }G^{\prime }}{(act)^{3(\omega +1)}}
\end{equation*}
that also reads:
\begin{equation}
G^{\prime }=\frac{a^{3(\omega +1)}dc^{5+3\omega }}{4\pi A_{\omega }}%
t^{3\omega }
\end{equation}
whose trivial integration is:
\begin{equation}
G=\frac{a^{3(\omega +1)}d}{4\pi }\frac{c^{5+3\omega }}{A_{\omega }}%
t^{3\omega +1}
\end{equation}
This result has already been obtained through the simplest technique, except
the numerical constant $\frac{a^{3(\omega +1)}d}{4\pi }.$

\paragraph{\textbf{Regardless of $div(T_{i}^{j})=0$}$.$}

If we do not take into account the condition $\ div(T_{ij})=0$ we shall
tackle the problem in the following way:
\begin{equation}
\rho ^{\prime }+3(\omega +1)\rho \frac{f^{\prime }}{f}+\frac{\Lambda
^{\prime }c^{4}}{8\pi G}+\rho \frac{G^{\prime }}{G}=0
\end{equation}
directly we obtain two pi-monomials: $\ \pi _{1}=\Lambda c^{2}t^{2}$ and $%
\pi _{2}=fc^{-1}t^{-1}$ that we replace into the equation:
\begin{equation}
\rho ^{\prime }+3(\omega +1)\rho \frac{1}{t}-\frac{dc^{2}}{4\pi t^{3}G}+\rho
\frac{G^{\prime }}{G}=0
\end{equation}
that we cannot simplify more. At this time, we note that with these
hypothesis we do not really get anything important.

\paragraph{\textbf{Considering the conditions $G=\protect\beta _{\protect%
\alpha }t$}$^{\mathbf{\protect\alpha }}$\textbf{\ and $div(T_{i}^{j})=0.$}}

If we take into account the hypothesis imposed by Kalligas et al (see \cite
{KA}) on the behavior of $G=Ct^{\alpha }$ with $\alpha \in \mathbb{R}$ the
dimensional treatment would be this: first, we define correctly the
(dimensional) constant that establishes the proportionality between $G$ and $%
t,$ $G=\beta _{\alpha }t^{\alpha }$ in such a way that $\beta _{\alpha }$
has the following dimensions: $\left[ \beta _{\alpha }\right]
=L^{3}M^{-1}T^{-2-\alpha }$ . As above we shall tackle the equation in the
same way:
\begin{equation}
\underset{A1}{\underbrace{\rho ^{\prime }+3(\omega +1)\rho \frac{f^{\prime }%
}{f}}}=-\underset{A2}{\underbrace{\left[ \frac{\Lambda ^{\prime }c^{4}}{8\pi
G}+\rho \frac{G^{\prime }}{G}\right] }}
\end{equation}
from $(A1)$ we obtain $\rho =A_{\omega }f^{-3(\omega +1)}$ and from ($A2)$
taking into account $\pi _{1}=\Lambda c^{2}t^{2}$ and $G=\beta _{\alpha
}t^{\alpha }$ we obtain the following relationship
\begin{equation}
\rho G^{\prime }=\frac{dc^{2}}{4\pi t^{3}}\qquad \rho =\frac{dc^{2}}{4\pi
\alpha \beta _{\alpha }}\frac{1}{t^{\alpha +2}}
\end{equation}
and therefore $f$
\begin{equation}
f=\left( \frac{A_{\omega }\beta _{\alpha }}{c^{2}}\right) ^{\frac{1}{%
3(\omega +1)}}t^{\frac{2+\alpha }{3(\omega +1)}}
\end{equation}
as we have seen, through the dimensional method the same results can be
obtain but in an easier way.

\paragraph{\textbf{Considering only the condition $G=\protect\beta _{\protect%
\alpha }t^{\protect\alpha }.$}}

Under this hypothesis we can also outline the problem without considering
the condition $div(T_{ij})=0.$ With these suppositions the equation to solve
is:
\begin{equation}
\rho ^{\prime }+3(\omega +1)\rho \frac{f^{\prime }}{f}+\rho \frac{\alpha }{t}%
+\frac{\Lambda ^{\prime }c^{4}}{8\pi \beta _{\alpha }t^{\alpha }}=0
\end{equation}
if furthermore, $\pi _{1}=\Lambda c^{2}t^{2}$ and $\pi _{2}=\rho \beta
_{\alpha }t^{\alpha +2}c^{-2}$ are taken into account then:
\begin{equation}
-\frac{(2+\alpha )gc^{2}}{\beta _{\alpha }t^{\alpha +3}}+3(\omega +1)\left(
\frac{gc^{2}}{\beta _{\alpha }t^{\alpha +2}}\right) \frac{f^{\prime }}{f}%
+\left( \frac{gc^{2}}{\beta _{\alpha }t^{\alpha +2}}\right) \frac{\alpha }{t}%
-\frac{dc^{2}}{4\pi \beta _{\alpha }t^{\alpha +3}}=0
\end{equation}
where $g$ and $d$ are numerical constants. Simplifying
\begin{equation*}
-\frac{(2+\alpha )g}{t}+3(\omega +1)gH+\frac{\alpha g}{t}-\frac{d}{4\pi t}=0
\end{equation*}
integrating
\begin{equation}
f=K_{\chi }t^{\chi }\qquad f=K_{\chi }t^{\frac{d+4\pi g(\alpha +1)}{3(\omega
+1)g}}
\end{equation}
where $\chi =\frac{d+4\pi g(\alpha +1)}{3(\omega +1)g}$ and $K_{\chi }$ is a
constant of proportionality with dimensions $\left[ K_{\chi }\right]
=LT^{-\chi }$ .

As we see, the recipe is always the same. First, consider the equation to be
integrated obtaining from it the highest number of pi-monomials, in order to
decrease the number of variables, in such a way that an easily integrable
equation is obtained.

\section{\textbf{Case $G$ and $\Lambda $ variable for a viscous fluid.}}

This problem was posed by Arbab (see \cite{AR}). The basic ingredients of
the model are:

The momentum-energy tensor defined by:
\begin{equation*}
T_{ij}=(\rho +p^{*})u_iu_j-pg_{ij}
\end{equation*}
where $\rho $ is the energy density and $p^{*}$ represents pressure $\left[
\rho \right] =\left[ p^{*}\right] $. The effect of viscosity is seen in:
\begin{equation}  \label{ee2}
p^{*}=p-3\xi H
\end{equation}
where: $p$ is the thermostatic pressure, $H=\left( f^{\prime }/f\right) $
and $\xi $ is the viscosity coefficient that follows the law:
\begin{equation}  \label{ee3}
\xi =k_\gamma \rho ^\gamma
\end{equation}
where $k_\gamma $ makes the equation be homogeneous i.e. it is a constant
with dimensions and where the constant $\gamma \in \left[ 0,1\right] $. And $%
p$ also verifies the next state equation:
\begin{equation}  \label{ee4}
p=\omega \rho \qquad \omega =const.
\end{equation}
where $\omega \in \left[ 0,1\right] $ (i.e. it is a pure number) so that the
momentum-energy tensor verifies the so-called energy conditions.

The field equations are:
\begin{equation}
2\frac{f\,^{\prime \prime }}{f\,}+\frac{(f\,^{\prime })^{2}}{f\,^{2}}=-\frac{%
8\pi G(t)}{c^{2}}p^{\ast }+c^{2}\Lambda (t)\ \   \label{a1}
\end{equation}
\begin{equation}
3\frac{(f\,^{\prime })^{2}}{f\,^{2}}=\frac{8\pi G(t)}{\,c^{2}}\rho
+c^{2}\Lambda (t)\qquad \quad \   \label{a2}
\end{equation}
deriving (\ref{a2}) and simplifying with (\ref{a1}) it yields
\begin{equation}
\rho ^{\prime }+3(\omega +1)\rho H-9k_{\gamma }\rho ^{\gamma }H^{2}+\frac{%
\Lambda ^{\prime }c^{4}}{8\pi G}+\rho \frac{G^{\prime }}{G}=0  \label{SI1}
\end{equation}
and at the moment we consider this equation.
\begin{equation}
div(T_{i}^{j})=0\text{ }\Leftrightarrow \rho ^{\prime }+3(\rho +p^{\ast })%
\frac{f^{\prime }}{f}=0  \label{a3}
\end{equation}
if we develop the equation (\ref{a3}) we get:
\begin{equation}
\rho ^{\prime }+3(\omega +1)\rho H-9k_{\gamma }\rho ^{\gamma }H^{2}=0
\label{e9}
\end{equation}

As in the case above, we shall study the model in two ways, one of them
analytic (non dimensional) and the other dimensional.

\subsection{\textbf{Non Dimensional Method}.}

In this section we will mainly follow Singh et al work (see \cite{SI}). If
we take the equation (\ref{SI1}) regrouped, we get:
\begin{equation}  \label{m1}
\underset{A1}{\underbrace{\rho ^{\prime }+3(\omega +1)\rho H-9k_\gamma \rho
^\gamma H^2}}=\underset{A2}{\underbrace{-\left[ \rho \frac{G^{\prime }}G+%
\frac{\Lambda ^{\prime }c^4}{8\pi G}\right] }}
\end{equation}
if take into account the conservation principle
\begin{equation}  \label{n5}
\rho ^{\prime }+3(\omega +1)\rho H-9k_\gamma \rho ^\gamma H^2=0
\end{equation}
then we solve this equation by solving the equation $A2$ in (\ref{m1}), in
such a way that the equation to be solved is now:
\begin{equation}  \label{n6}
\left[ \rho \frac{G^{\prime }}G+\frac{\Lambda ^{\prime }c^4}{8\pi G}\right]
=0
\end{equation}
this equation is tried to be solved like this (see \cite{SI}). $\Lambda =
\frac{3\beta H^2}{c^2}$ \textbf{is defined }(hypothesis by Arbab (see \cite
{AR}) as well as by Singh et al (see \cite{SI}), condition that as we shall
see, is not necessary to impose in the solution through D.A.) and from the
equation (\ref{a2}) the following relationship is obtained: $8\pi G\rho
=3(1-\beta )H^2$. Therefore if all the equalities are replaced in the
equation (\ref{n6}) it yields:
\begin{equation}  \label{n7}
\frac 2{(1-\beta )}\frac{H^{\prime }}H=\frac{\rho ^{\prime }}\rho
\end{equation}
which is easily integrated.
\begin{equation}  \label{n8}
H=C_1\rho ^{1/d}\qquad d=\frac 2{(1-\beta )}
\end{equation}
we get to the equation (\ref{n5}) with all these results
\begin{equation*}
\rho ^{\prime }+3(\omega +1)\rho H-9k_\gamma \rho ^\gamma H^2=0
\end{equation*}
we arrive to the next equation:
\begin{equation}  \label{l1}
\rho ^{\prime }+3C_1(\omega +1)\rho ^{\frac{d+1}d}-9C_1^2k_\gamma \rho ^{%
\frac{d\gamma +2}d}=0
\end{equation}
which has got a particular solution in the case $\gamma =d^{-1}$ obtaining:
\begin{equation*}
\rho (t)=\dfrac 1{\left( a_0t\right) ^d}\qquad /\text{ }a_0=\left(
3C_1(\omega +1)-9k_\gamma C_1^2\right) d^{-1}
\end{equation*}
and obtaining from it:
\begin{equation*}
f(t)=C_2t^{\dfrac 1{\left( 3(\omega +1)-3k_\gamma C_1\right) (1-\gamma )}}
\end{equation*}
This is the most developed solution reached by Singh \ et al (see \cite{SI})
which is slightly different from the one by Arbab (see \cite{AR}).

\subsection{\textbf{Dimensional Method.}}

We shall explore two dimensional methods in this section. The first one,
probably the simplest one, has the inconvenience of having to depend on
Einstein criterion(see \cite{EIN} and Barenblatt \cite{BAR}), while the
second one is more powerful and more elaborated. We shall finish showing an
equation obtained without having to impose the condition $div(T_{i}^{j})=0.$

\subsubsection{\textbf{Naive Method.}}

We summarized here the method addressing to the reference (\cite{T3}) to
know more deeply the followed method. In this model the set of quantities
and constant $\frak{M}$ remain reduced to $\frak{M=M}(t,c,A_{\omega
},k_{\gamma })$ while the dimensional base continues being $\frak{B}%
=\{L,M,T,\theta \}.$ The followed dimensional method leads us to the
obtainment of two dimensionless $\pi -$monomials $\pi _{1}=\varphi (\pi
_{2}) $ where $\varphi $ represents a unknown function. To arrive to a
satisfactory solution i.e. to get rid of such function $\varphi ,$ we shall
have to take into account the criteria of Einstein or Barenblatt. With this
criterion we obtain the same results as the already existing ones in the
literature, but let us say, in a trivial way (for more details see \cite{T3}%
).

\subsubsection{\textbf{Not so simple Method}}

In this section we shall combine dimensional techniques with standard
techniques of ODEs integration. With the dimensional method we shall obtain
dimensionless monomials, which will be replaced in the equations. Thus, the
number of variables will be reduced in such way that the resulting equation
is integrable in a trivial way. We study two cases, the first one in which
we consider \textbf{$div(T_{i}^{j})=0,$} while in the other, as we shall
see, such hypothesis is not needed.

\paragraph{\textbf{Considering the condition $div(T_{i}^{j})=0.$}}

In this case we shall pay attention into the equation:
\begin{equation*}
\rho ^{\prime }+3(\omega +1)\rho H-9k_{\gamma }\rho ^{\gamma }H^{2}+\rho
\frac{G^{\prime }}{G}+\frac{\Lambda ^{\prime }c^{4}}{8\pi G}=0
\end{equation*}
taking into account the relationship $div(T_{i}^{j})=0$ The following
equality is brought up:
\begin{equation}
\underset{A1}{\underbrace{\rho ^{\prime }+3(\omega +1)\rho H-9k_{\gamma
}\rho ^{\gamma }H^{2}}}=\underset{A2}{\underbrace{-\left[ \rho \frac{%
G^{\prime }}{G}+\frac{\Lambda ^{\prime }c^{4}}{8\pi G}\right] }}
\label{SILVI}
\end{equation}

The idea is the following: By using D.A. we obtain two $\pi -$monomials,
which are replaced in the equation, obtaining a huge simplification of it.
On the other hand we integrate ($A1)$ and $(A2),$ solving the problem
completely in this way, now without Barenblatt`s criterion. Let see. The
monomials obtained are: $\ \pi _{1}=\rho k_{\gamma }^{\frac{-1}{1-\gamma }%
}t^{\frac{-1}{\gamma -1}}$ and $\pi _{2}=\Lambda c^{2}t^{2}$ \ i.e.
\begin{equation*}
\rho =ak_{\gamma }^{\frac{1}{1-\gamma }}t^{\frac{1}{\gamma -1}}\qquad
\Lambda =\dfrac{d}{c^{2}t^{2}}
\end{equation*}
where $a$ and $d$ are numerical constants. In a generic way the solution is
of the following way: $\rho =ak_{\gamma }^{\frac{1}{1-\gamma }}t^{\frac{1}{%
\gamma -1}}$ \ if we define $b=\frac{1}{1-\gamma }$ then $\rho =ak_{\gamma
}^{b}t^{-b}$ where $a=const.\in \mathbb{R}$ then $\rho ^{\prime
}=-bak_{\gamma }^{b}t^{-b-1}$ (paying attention only to the term $(A1)$ of
the equation) it yields:
\begin{equation}
-bak_{\gamma }^{b}t^{-b-1}+3(\omega +1)ak_{\gamma }^{b}t^{-b}H-9k_{\gamma
}\left( ak_{\gamma }^{b}t^{-b}\right) ^{\gamma }H^{2}=0
\end{equation}
that simplifying it is reduced to:
\begin{equation}
9a^{\left( \gamma -1\right) }\left( f^{\prime }\right)
^{2}-3wt^{-1}ff^{\prime }+bt^{-2}f^{2}=0
\end{equation}
\begin{equation}
f^{\prime }=\frac{f}{t}\left[ \frac{1}{6a^{\gamma -1}}\left( w\pm
(w^{2}-4ba^{\gamma -1})^{\frac{1}{2}}\right) \right]
\end{equation}
where $w=(\omega +1),$ if define
\begin{equation}
D=\left[ \frac{1}{6a^{\gamma -1}}\left( w\pm (w^{2}-4ba^{\gamma -1})^{\frac{1%
}{2}}\right) \right]
\end{equation}
then, the solution has the following form:
\begin{equation}
f=lBt^{D}
\end{equation}
where $l$ is a certain numerical constant and $B$ is a integration constant
with dimensions, that can be identified with our result by making $%
B=A_{\omega }k_{\gamma }$.

Now we shall solve the other term of the equation (the $A2)$. the equation ($%
\left[ \rho \frac{G^{\prime }}{G}+\frac{\Lambda ^{\prime }c^{4}}{8\pi G}%
\right] =0$ (\ref{n6})) can be solved in a trivial way if we follow the next
results. If we replace the monomials $\ \pi _{1}=\rho k_{\gamma }^{\frac{-1}{%
1-\gamma }}t^{\frac{-1}{\gamma -1}}$ \ and $\pi _{2}=\Lambda c^{2}t^{2}$ \
in such equation the integration of it becomes trivial:
\begin{equation*}
ak_{\gamma }^{\frac{1}{1-\gamma }}t^{\frac{1}{\gamma -1}}\left( \frac{%
G^{\prime }}{G}\right) -\frac{dc^{2}}{4\pi Gt^{3}}=0
\end{equation*}
\begin{equation}
G^{\prime }=\frac{dc^{2}}{a4\pi k_{\gamma }^{b}}t^{b-3}\Longrightarrow G(t)=g%
\frac{dc^{2}}{a4\pi k_{\gamma }^{b}}t^{b-2}
\end{equation}
where $a,d$ and $g\in \mathbb{R}$ (they are pure numbers). We can also
observe that this integral needs not be solved since a more careful analysis
on the number of $\pi -$monomials that we can obtain from the equation,
leads us to obtain a solution of the type:
\begin{equation*}
G=G(k_{\gamma },c,t)
\end{equation*}
which brings us to:
\begin{equation}
G(t)=gk_{\gamma }^{-b}c^{2}t^{b-2}
\end{equation}
This method, as we have seen, is more prepared (strived) and the solution,
therefore, finer though coincident with the previous one.

\paragraph{\textbf{Case in which $div(T_{i}^{j})=0$ is not considered.}}

Let see now how we can tackle (broach) this problem from the D.A. point of
view, without imposing the condition $div(T_{i}^{j})=0$. The base $\frak{B}$
as before, is still $\frak{B}=\left\{ L,M,T\right\} $ while the fundamental
set of quantities and constant this time is $\frak{M=M}(t,c,k_{\gamma })$,
with these data we can obtain the following monomials from the equation\
\begin{equation}
\rho ^{\prime }+3(\omega +1)\rho H-9k_{\gamma }\rho ^{\gamma }H^{2}+\rho
\frac{G^{\prime }}{G}+\frac{\Lambda ^{\prime }c^{4}}{8\pi G}=0  \label{BER1}
\end{equation}
considering that:
\begin{equation}
\rho =ak_{\gamma }^{\frac{1}{1-\gamma }}t^{\frac{1}{\gamma -1}}\qquad
\Lambda =\dfrac{d}{c^{2}t^{2}}  \label{BER2}
\end{equation}
this two monomials are replaced into the equation, which is quite
simplified:
\begin{equation}
-bak_{\gamma }^{b}t^{-b-1}+3(\omega +1)ak_{\gamma }^{b}t^{-b}H-9k_{\gamma
}\left( ak_{\gamma }^{b}t^{-b}\right) ^{\gamma }H^{2}+ak_{\gamma }^{b}t^{-b}%
\frac{G^{\prime }}{G}-\frac{dc^{2}}{4\pi Gt^{3}}=0  \label{BER3}
\end{equation}
simplifying this equation, it yields:
\begin{equation}
-9a^{\left( \gamma -1\right) }tH^{2}+3wH-bt^{-1}+\frac{G^{\prime }}{G}-\frac{%
dc^{2}}{4\pi ak_{\gamma }^{b}}\dfrac{t^{b-3}}{G}=0  \label{BER4}
\end{equation}
that along with the field equations (\ref{a1}) and (\ref{a2}) carry us to
the next set of equations. For example we see:
\begin{equation*}
3H^{2}=a\frac{8\pi G}{c^{2}}k_{\gamma }^{b}t^{-b}+\frac{d}{t^{2}}
\end{equation*}
that we replaced into the equation that we are treating (dealing with),
resulting:
\begin{equation*}
-bt^{-1}+3w\left( \frac{a8\pi k_{\gamma }^{b}}{3c^{2}}Gt^{-b}+\frac{d}{3t^{2}%
}\right) ^{\frac{1}{2}}-
\end{equation*}
\begin{equation*}
-9a^{\left( \gamma -1\right) }\left( \frac{a8\pi k_{\gamma }^{b}}{3c^{2}}%
Gt^{-b}+\frac{d}{3t^{2}}\right) t+\frac{G^{\prime }}{G}-\frac{dc^{2}}{4\pi
ak_{\gamma }^{b}}\dfrac{t^{b-3}}{G}=0
\end{equation*}
that solving it results:
\begin{equation}
G=gk_{\gamma }^{-b}c^{2}t^{b-2}
\end{equation}
where $g\in \mathbb{R}$ represents a numerical constant. We finally observe
that as in the previous section we could have taken into account the three
monomials obtained from the equation i.e.
\begin{equation*}
\rho =ak_{\gamma }^{b}t^{-b}\qquad \Lambda =\dfrac{d}{c^{2}t^{2}}\qquad G=g%
\frac{c^{2}t^{b-2}}{k_{\gamma }^{b}}
\end{equation*}
replace them into the equation
\begin{equation*}
\rho ^{\prime }+3(\omega +1)\rho H-9k_{\gamma }\rho ^{\gamma }H^{2}+\rho
\frac{G^{\prime }}{G}+\frac{\Lambda ^{\prime }c^{4}}{8\pi G}=0
\end{equation*}
and calculate $f,$ arriving at the same solution obtained in the section
above i.e.
\begin{equation*}
f=lBt^{D}
\end{equation*}
We have proved that it is not necessary to impose the condition $%
div(T_{i}^{j})=0$ since, in this case, the same solution is obtained as
imposing it.

\section{\textbf{Case $G,c$ and $\Lambda $ variable and viscous fluid.}}

This model has been developed exclusively through dimensional techniques,
therefore, at this time no other solution is known, except the one showed in
section $\frak{4.2}$. We can consider that this model is a natural
generalization of the previous one (see \cite{T4}).

Following the Alberch and Magueijo work and therefore all the assumptions
considered there, we arrive to the requirement that the standard equations
still retain their form but with $G(t),c(t)$ and $\Lambda (t)$ varying (see
\cite{AM}).

The field equations that describe our viscous model are (in this section we
shall take into account the hypothesis made in the above section):
\begin{equation}
2\frac{f\,^{\prime \prime }}{f\,}+\frac{(f\,^{\prime })^{2}}{f\,^{2}}=-\frac{%
8\pi G(t)}{c^{2}(t)}p^{\ast }+c^{2}(t)\Lambda (t)\ \   \label{MI1}
\end{equation}
\begin{equation}
3\frac{(f\,^{\prime })^{2}}{f\,^{2}}=\frac{8\pi G(t)}{\,c^{2}(t)}\rho
+c^{2}(t)\Lambda (t)\qquad \quad \   \label{MI2}
\end{equation}
developing equation (\ref{eg2}) i.e. calculating $\left( \frac{8\pi G}{c^{4}}%
T_{i}^{j}+\Lambda \delta _{i}^{j}\right) _{;j}$ in this case it yields:
\begin{equation}
\rho ^{\prime }+3(\omega +1)\rho H-9k_{\gamma }\rho ^{\gamma }H^{2}+\frac{%
\Lambda ^{\prime }c^{4}}{8\pi G}+\rho \frac{G^{\prime }}{G}-4\frac{c^{\prime
}}{c}\rho =0  \label{MI3}
\end{equation}
besides the so-called conservation principle for the energy-momentum tensor
is taken into account
\begin{equation}
div(T_{i}^{j})=0\text{ }\Leftrightarrow \rho ^{\prime }+3(\rho +p^{\ast })%
\frac{f^{\prime }}{f}=0  \label{MI4}
\end{equation}
Now, if we develop the equation (\ref{MI4}) it is obtained:
\begin{equation}
\rho ^{\prime }+3(\omega +1)\rho H-9k_{\gamma }\rho ^{\gamma }H^{2}=0
\label{MI5}
\end{equation}
Under these circumstances the equation to be treated is:
\begin{equation}
\rho ^{\prime }+3(\omega +1)\rho H-9k_{\gamma }\rho ^{\gamma }H^{2}=-\left[
\frac{\Lambda ^{\prime }c^{4}}{8\pi G}+\rho \frac{G^{\prime }}{G}-4\frac{%
c^{\prime }}{c}\rho \right]   \label{m2}
\end{equation}

We cannot play the same trick as before since in this case the ``constant'' $%
c\mapsto c(t)$ varies and therefore the equation are much more complicated.
We shall show here two methods: the naive and the elaborated one (not so
simple).

\subsection{\textbf{Naive Method.}}

Since in this case we are considering the ``constant'' $c$ as variable, the
set of governing quantities remains reduced to $\frak{M=M}(t,k_{\gamma
},A_{\omega })$ while the dimensional base is still $\frak{B}=(L,M,T)$. The
obtained solutions in this case are perfectly definite since a single $\pi -$%
monomial is obtained. To see more details about this technique we address to
reference \cite{T4}.

\subsection{\textbf{Not so simple Method.}}

In this occasion the equation to solve is (considering the condition $%
div(T_{ij})=0$, as we cannot get rid of it):
\begin{equation}
\underset{A1}{\underbrace{\rho ^{\prime }+3(\omega +1)\rho H-9k_{\gamma
}\rho ^{\gamma }H^{2}}}=\underset{A2}{\underbrace{-\left[ \frac{\Lambda
^{\prime }c^{4}}{8\pi G}+\rho \frac{G^{\prime }}{G}-4\frac{c^{\prime }}{c}%
\rho \right] }}  \label{SIL1}
\end{equation}
This equation, is very similar to equation (\ref{SILVI}). We shall follow in
this paragraph the tactics developed in section $\frak{3.2.2}$. On the one
hand, we already know how to tackle the left side of equation (\ref{SIL1})
i.e. the $A1$ side. If we take into account these results and \textbf{%
imposing} the following \textbf{condition} $\Lambda =\dfrac{d}{c^{2}(t)t^{2}}
$ (that might be unfounded) we can simplify this equation:
\begin{equation*}
-bt^{-1}+3wH-9a^{\left( \gamma -1\right) }tH^{2}+\frac{G^{\prime }}{G}-\frac{%
2dc\left[ c^{\prime }t+c\right] }{8\pi ak_{\gamma }^{b}}\dfrac{t^{b-3}}{G}-4%
\dfrac{c^{\prime }}{c}=0
\end{equation*}
with the usual notation that we are following. From one of the field
equations expression for $G$ has been \ obtained.
\begin{equation*}
G=\dfrac{c^{2}}{8\pi ak_{\gamma }^{b}t^{-b}}\left[ 3H^{2}-\dfrac{d}{t^{2}}%
\right]
\end{equation*}
also $f$ is a previous result from last section:
\begin{equation*}
f=lBt^{D}
\end{equation*}
We should not forget that ($A1$) from equation (\ref{SIL1}) is the same as
the one studied in section $\frak{3.2.2.}$ (\ref{SILVI}) and $\rho $ makes
no difference. With this expressions we continue our simplification. Since $%
H^{2}=\dfrac{D^{2}}{t^{2}}$ then
\begin{equation}
G=\dfrac{(3D^{2}-d)c^{2}}{8\pi ak_{\gamma }^{b}}t^{b-2}
\end{equation}
We observe again that $c=c(t)$.
\begin{equation*}
G^{\prime }=\dfrac{2(3D^{2}-d)cc^{\prime }}{8\pi ak_{\gamma }^{b}}%
t^{b-2}+(b-2)\dfrac{(3D^{2}-d)c^{2}}{8\pi ak_{\gamma }^{b}}t^{b-3}
\end{equation*}
\begin{equation*}
\dfrac{G^{\prime }}{G}=2\dfrac{c^{\prime }}{c}+\dfrac{b-2}{t}
\end{equation*}
with this results we go on to simplify the term $(A2)$ in equation (\ref
{SIL1}) obtaining
\begin{equation*}
\frac{G^{\prime }}{G}+\frac{2dc\left[ c^{\prime }t+c\right] }{8\pi
ak_{\gamma }^{b}}\dfrac{t^{b-3}}{G}-4\dfrac{c^{\prime }}{c}=0
\end{equation*}
\begin{equation*}
2\dfrac{c^{\prime }}{c}+\dfrac{b-2}{t}+\left[ \frac{-2d}{(3D^{2}-d)}\right]
\left( \dfrac{c^{\prime }}{c}+\dfrac{1}{t}\right) -4\dfrac{c^{\prime }}{c}=0
\end{equation*}
\begin{equation*}
\dfrac{c^{\prime }}{c}=\left( \dfrac{b}{2}-1-\dfrac{bd}{6D^{2}}\right)
\dfrac{1}{t}
\end{equation*}
\begin{equation}
c=K_{\chi }t^{\chi }
\end{equation}
where $\chi $$=\left( \dfrac{b}{2}-1-\dfrac{bd}{6D^{2}}\right) $, and $%
K_{\chi }$ is the proportionality constant that relates $c$ with $t$ with
dimensions $\left[ K_{\chi }\right] =LT^{-1-\chi }$. From this result the
expression of $G$ is obtained:
\begin{equation}
G=\dfrac{(3D^{2}-d)K_{\chi }^{2}}{8\pi ak_{\gamma }^{b}}t^{b-2+2\chi }
\end{equation}
With this solution, it is observed for the case $\gamma =\frac{1}{2}$ what
corresponds to $b=2$
\begin{equation}
\frac{G}{c^{2}}=\frac{t^{b-2+2\chi }}{t^{2\chi }}=const.
\end{equation}
i.e. \textbf{the covariance principle is verified}. In particular if $b=2$
then $\chi $$=\left( \dfrac{-d}{3D^{2}}\right) $.

While the cosmological ``constant'' yields $\left( \Lambda =\frac{d}{%
c^{2}(t)t^{2}}\right) :$%
\begin{equation}
\Lambda =\frac{d}{c^{2}(t)t^{2}}=\frac{d}{K_{\chi }^{2}t^{2\chi +2}}
\end{equation}

In this way the remaining quantities are calculated.

\section{\textbf{Case of $G,c$ and $\Lambda $ variable with adiabatic matter
creation.}}

The momentum-energy tensor is defined by:
\begin{equation}
T_{ij}=(\rho +p^{\ast }+p_{c})u_{i}u_{j}-(p^{\ast }+p_{c})g_{ij}
\end{equation}
where $\rho $ is the energy density and $p^{\ast }$ represents pressure $%
\left[ \rho \right] =\left[ p^{\ast }\right] $. The effect of the viscosity
is given by:
\begin{equation*}
p^{\ast }=p-3\xi H
\end{equation*}
where: $p$ is the thermostatic pressure, $H=\left( f^{\prime }/f\right) $
and $\xi $ is the viscosity coefficient that follows the law:
\begin{equation}
\xi =k_{\gamma }\rho ^{\gamma }
\end{equation}
The field equations are as it follows:
\begin{equation}
2\frac{f\,^{\prime \prime }}{f\,}+\frac{(f\,^{\prime })^{2}}{f\,^{2}}=-\frac{%
8\pi G(t)}{c^{2}(t)}(p^{\ast }+p_{c})+c^{2}(t)\Lambda (t)\ \
\end{equation}
\begin{equation}
\frac{(f\,^{\prime })^{2}}{f\,^{2}}=\frac{8\pi G(t)}{3\,c^{2}(t)}\rho
+c^{2}(t)\Lambda (t)\qquad \quad \
\end{equation}
\begin{equation}
n^{\prime }+3nH-\psi =0
\end{equation}
where $n$ measures the particles number density, $\psi $ is the function
that measures the matter creation, $H=f^{\prime }/f$ represents the Hubble
parameter ($f$ is the scale factor that appears in the metric), $p$ is the
thermostatic pressure, $\rho $ is energy density and $p_{c}$ is the pressure
that generates the matter creation.

The creation pressure $p_{c}$ depends on the function $\psi $. For adiabatic
matter creation this pressure takes the following form:
\begin{equation}
p_{c}=-\left[ \frac{\rho +p}{3nH}\psi \right]  \label{w2}
\end{equation}
The state equation that we next use is the well-known expression
\begin{equation}
p=\omega \rho  \label{w3}
\end{equation}
where $\omega =const.$ $\omega \in \left[ 0,1\right] $ physically realistic
equations, making in this way that the energy-momentum tensor $T_{ij}$
verifies the energy conditions.

We need to know the exact form of the function $\psi $ , which is determined
from a more fundamental theory that involves quantum processes. We assume
that this function follows the law:
\begin{equation}
\psi =3\beta nH  \label{w5}
\end{equation}
where $\beta $ is a dimensionless constant (if $\beta =0$ then there is no
matter creation since $\psi =0)$ presumably given by models of particles
physics of matter creation.

The generalized principle of conservation brings us to:
\begin{equation}
\rho ^{\prime }+3(\omega +1)(1-\beta )\rho H-9k_{\gamma }\rho ^{\gamma
}H^{2}=-\left[ \frac{\Lambda ^{\prime }c^{4}}{8\pi G}+\rho \frac{G^{\prime }%
}{G}-4\frac{c^{\prime }}{c}\rho \right]
\end{equation}
The conservation principle for the momentum-energy tensor is expressed
through the following law:
\begin{equation*}
\rho ^{\prime }+3(\rho +p+p_{c}-3\xi H)H=0
\end{equation*}
\begin{equation}
\rho ^{\prime }+3(\omega +1)(1-\beta )\rho H-9k_{\gamma }\rho ^{\gamma
}H^{2}=0
\end{equation}
This equation has been solved in the above section since the only difference
between this one and the one exposed there is the term $(1-\beta )$ and the
solution will be very similar. The dimensional solution can be found in the
reference (\cite{T5}).

\section{\textbf{Conclusions.}}

The purpose of this work has been to show how to solve this type of models
through dimensional techniques. We have been able to prove that this technic
is applied in a comparatively easy way enabling us to obtain almost trivial
solutions and allowing to avoid hypotheses needed when using other
methods.\bigskip

\begin{acknowledgement}
We wish to express our gratefulness to Prof. Manuel Casta\~{n}s for his help
and critical commentaries that they have helped the successful attainment of
this work and to Javier Aceves for his collaboration in the translation into
English.
\end{acknowledgement}

\end{document}